\pdfoutput=1

\documentclass{ws-ijmpa}

\usepackage[super,compress]{cite}
\usepackage{graphicx}

\def\p{\partial}
\def\dfrac#1#2{{\displaystyle\frac{#1}{#2}}}
\def\stTD#1#2{\hbox to 0em{\mathsurround=0em $\stackrel{#1}{\makebox[0pt]{} #2}$\hss} \phantom{#2}}\def\stscript#1#2{\hbox to 0em{\mathsurround=0em ${\scriptstyle\stackrel{#1}{\makebox[0pt]{} #2}}$\hss} \phantom{#2}}\def\stscriptscript#1#2{\hbox to 0em{\mathsurround=0em ${\scriptscriptstyle\stackrel{#1}{\makebox[0pt]{} #2}}$\hss} \phantom{#2}}
\def\comb#1#2#3{{\mathsurround 0pt\hbox to 0pt {\hspace*{#3}\raisebox{#2}{${#1}$}\hss}}}
\def\combs#1#2#3{{\mathsurround 0pt\hbox to 0pt {\hspace*{#3}\raisebox{#2}{${\scriptstyle #1}$}\hss}}}
\def\combss#1#2#3{{\mathsurround 0pt\hbox to 0pt {\hspace*{#3}\raisebox{#2}{${\scriptscriptstyle #1}$}\hss}}}
\def\df{\mathrm{d}}
\def\Vols{V}
\def\Energy{\mathbb{E}}
\def\EMV{\mathbb{P}}
\def\AMV{\mathbb{J}}
\def\Act{\mathcal{A}}
\def\Vol{\overline{V}}
\def\metr{\mathfrak{m}}
\def\Fem{F}
\def\eqdef{\doteqdot}
\def\Ae{A}
\def\Surc{\sigma}
\def\Fo{\mathbb{F}}

\def\xp#1{\comb{\cdot}{-0.9ex}{0.3ex}{#1}}
\def\xc{\bar{x}}
\def\metrEff{\mathchoice{\combs{\sim}{1ex}{0.2ex}\mathfrak{m}}{\combs{\sim}{1ex}{0.2ex}\mathfrak{m}}{\combss{\sim}{0.66ex}{0.05ex}\mathfrak{m}}{}{}}

\begin{document}

\markboth{A.~A.~Chernitskii}{Fundamental interactions and quantum behaviour in unified field theory}

%
\catchline{}{}{}{}{}
%

\title{Fundamental interactions and quantum behaviour in unified field theory}

\author{Alexander A. Chernitskii}

\address{Department of Mathematics\\ St. Petersburg State Chemical Pharmaceutical University\\Prof. Popov str. 14, St. Petersburg, 197022, Russia\\[0.5ex]
and\\[0.5ex]
A. Friedmann Laboratory for Theoretical Physics\\St. Petersburg, Russia\\AAChernitskii@mail.ru}

\maketitle

\begin{history}
\received{Day Month Year}
\revised{Day Month Year}
\end{history}

\begin{abstract}
The concept of unified field theory is discussed.
Two nonlinear field models with world volume type action are considered, namely extremal space-time film model
and  Born -- Infeld nonlinear electrodynamics.
The natural appearance of two long-range fundamental interactions, electromagnetism and gravitation, in these field models is discussed.
The quantum behaviour of the interacting solitons-particles is considered.
The concept of quasi-bounding quantization in nonlinear field models is introduced.

\end{abstract}

\ccode{PACS numbers: 04.50.-h, 12.10.-g}

\section{Introduction}
\label{introd}
The concept of unified field theory is known for a long time.
According to this concept we consider a fundamental nonlinear field model with the following two relations to the material world:
\begin{itemize}
\item The elementary particles must be represented as space-localized solutions of this model.
\item All interactions between the particles must appear naturally as the manifestation of the nonlinearity of the model.
\end{itemize}
It is well known that the creator of relativity theory A. Einstein was a proponent of this concept.\cite{Einstein1953aE}
Also many famous researchers worked in the framework of this concept.

But we see a relatively weak progress in this direction of investigation. This is connected with the
appropriate extraordinary mathematical difficulties.
Recent defined advancement in the part of finding of exact solutions for a nonlinear field model\cite{Chernitskii2018a} can give a base for
the physical accordance between this theory and reality.

\section{Unified Field Models and Solitons-Particles}
\label{worldva}
There are two basic questions concerning the choice of an unified field model:
the tensor character of the field
and
the invariance group of the model equations.
Also we must define a space dimension which will be considered.

Here we consider two generally invariant models in four-dimensional space-time,
namely the scalar model of extremal space-time film\cite{Chernitskii2018a,Chernitskii2016a} and
the vector model of nonlinear Born--Infeld electrodynamics\cite{Chernitskii1999,Chernitskii2004a}.
The extremal space-time film\cite{Chernitskii2018a} model is the relativistic generalization
of the minimal surface or minimal thin film one in three-dimensional space. It is sometimes
called Born--Infeld type scalar field model\cite{BarbChern1967-1e}.

Both models are formulated in the form of extremal action principle, where the action has the following world volume form:
\begin{equation}
\label{35135655}
\Act  =\int_{\Vol}\!\sqrt{|\mathfrak{M}|}\;(\mathrm{d}x)^{4}
\;.
\end{equation}
Here $\Vol$ is a space-time volume, $\mathfrak{M} \eqdef \det(\mathfrak{M}_{\mu\nu})$ is determinant of a world tensor field
$\mathfrak{M}_{\mu\nu}$,
 $(\mathrm{d}x)^{4} \eqdef \mathrm{d}x^0\mathrm{d}x^1\mathrm{d}x^2\mathrm{d}x^3$.
\begin{subequations}\label{669446731}
We consider the
field
$\mathfrak{M}_{\mu\nu}$ in the following
forms:
\begin{align}
\label{669707871}
  \mathfrak{M}_{\mu\nu} &=
  \metr_{\mu\nu} + \chi^2\,\Phi_{\mu}\,\Phi_{\nu}   \;,\quad \Phi_{\mu} \eqdef \frac{\p \Phi}{\p x^\mu}
\;,\\
\label{669707872}
\mathfrak{M}_{\mu\nu} &= \metr_{\mu\nu} + \chi^2\,\Fem_{\mu\nu}
\;,\quad
\Fem_{\mu\nu} \eqdef \frac{\p \Ae_\nu}{\p x^\mu} {}-{}
\frac{\p \Ae_\mu}{\p  x^\nu}
\;,
\end{align}
where $\Phi$ is a scalar field, $\Ae_\mu$ are components of electromagnetic potential vector.
\end{subequations}

We use the term ``soliton'' in a wide sense that is a space-localized solution of a nonlinear field model.

The simplest static spherically symmetric solutions of the field models under consideration are known.\cite{Chernitskii2017a,Chernitskii1999}
A time-periodic soliton solution with a static part can represent a real massive particle at rest.
But such solutions are not known at present.

Recently the time-periodic solutions propagating with the speed of light was obtained for the space-time film model.
A subclass of these solutions has a correspondence with photons.\cite{Chernitskii2018a,Chernitskii2016a}

\section{Fundamental Interactions}
\label{fundint}

We know that a linear field model is characterized by superposition property.
According to this property a linear combination of solutions is also a solution of the linear model.
To obtain an interaction effect in a linear model we must introduce additional conditions.
In this way Coulomb--Lorentz force is introduced in the linear classical electrodynamics.

The nonlinearity of the model violates the superposition property for its solutions.
This violation must be interpreted as an interaction between solitons considered as particles.
Thus the interaction effect is a native property of such models.

First of all the realistic unified field model must provide two long-range interactions between the particles, namely electromagnetism and gravitation.
Both extremal space-time film model and Born--Infeld electrodynamics provide these interactions.%
\cite{Chernitskii1999,Chernitskii2004a,Chernitskii2012be,Chernitskii2017a,Chernitskii2016b}

Let us write the following integral conservation law:\cite{Chernitskii2016b}
\begin{equation}
\label{509874761}
\dfrac{\df \EMV^\mu_{\Vols}}{\df \xc^0} = \Fo^\mu_{\Surc}
\;,
\end{equation}
where $\EMV^\mu_{\Vols}$ is a momentum of the field in the three-dimensional volume $\Vols$,
$\Fo^\mu_{\Surc}$ is the integral force which is a surface integral over the border $\Surc$ of the volume $\Vols$.

Application (\ref{509874761})
to the long-range soliton interaction gives the appropriate soliton dynamic equation in some approximation.
In this case the volume $\Vols$ is a localization region of the soliton.
The force $\Fo^\mu_{\Surc}$ is caused by a small field of the distant solitons.
Thus this soliton interaction can be called the force interaction of solitons.\cite{Chernitskii2016b}
It must be emphasized that the obtained dynamical law (\ref{509874761}) is model independent.

The considered force interaction applying to a relativistically invariant field model gives
the Newtonian dynamic law with the Coulomb -- Lorentz force.%
\cite{Chernitskii1999,Chernitskii2012be,Chernitskii2017a}
Thus we have the electromagnetic interaction between soliton-particles in this case.

It should be noted that the mass of the interacting soliton-particle appears as its full energy in intrinsic
coordinate system.
Thus we have here the substantiation for Einsteinian postulate of equivalence between mass and energy.

The tensor character of the electromagnetic field is caused by the tensor character of the integral force integrand.\cite{Chernitskii2017a}

Gravitational interaction between solitons appears as a consequence of consideration of notable characteristic equations for the field models
under investigation.
For the long-range interacting solitons with fast-oscillating parts we have the following dispersion relation:\cite{Chernitskii2016b}\vspace{-2ex}
\begin{equation}
\label{396089081}
\left|\metrEff^{\mu\nu}\,k_{\mu}\,k_{\nu}\right|  =  \xp{\omega}^2
\;,
\end{equation}
where $k_{\mu}$ is a wave vector of the soliton,  $\metrEff^{\mu\nu}$ is an effective metric which is caused by a field of distant solitons,
 $\xp{\omega}$ is an angular frequency of the soliton time-periodic part in its intrinsic inertial coordinate system.

In this case the trajectory equation of the interacting soliton is the same that we have in general relativity theory, that is the
appropriate geodesic equation.
The type of soliton long-range interaction caused by the effective metric can be called the metrical one.\cite{Chernitskii2016b}

It can be shown,\cite{Chernitskii2016b,Chernitskii2012be}
to consider the metrical soliton interaction as the real gravitation we must take into account the wave background
field of all soliton-particles in the universe.
Thus the wave or quantum properties of soliton-particles is essential for the gravitation interpretation of
metrical interaction.

\section{Quantum Behaviour}
\label{quantbeh}
The soliton-particle with time-periodic part in its intrinsic coordinate system manifests both corpuscular and  wave properties in a coordinate system where the
soliton is moving. And the quantum behaviour of the soliton is caused by its wave properties.

Let us consider the effect of quasi-bounding quantization in nonlinear field theory.
The bounding quantization is a discretization for a wave frequency because of the presence of boundaries.
This quantization effect is well known for linear field models and appears in resonators.

The quasi-bounding quantization appears in nonlinear field theory because of interaction effect.
One solution can create a quasi boundary for an additional small field.
The appropriate problem has a form of the wave propagation one in inhomogeneous medium.
In particular, this quantization effect appears in a scattering problem on the static spherically symmetric
solution in
 nonlinear electrodynamics.\cite{Chernitskii2006b}

Also we have the correspondence between first-order twisted lightlike solitons of extremal space-time film and photons.\cite{Chernitskii2018a,Chernitskii2016a}
In high-frequency approximation ($\omega\to\infty$) we have the following notable relations between energy $\Energy$, momentum $\EMV$, and angular momentum $\AMV$ of the twisted lightlike soliton:\cite{Chernitskii2018a}
\begin{equation}
\label{832897461}
\Energy  = \EMV  = \frac{\omega}{m}\,\AMV
\;.
\end{equation}
Thus for the first-order twisted lightlike solitons ($m=1$), relation (\ref{832897461}) becomes appropriate to photons.

A consideration of the ideal gas of the first-order twisted lightlike solitons in finite volume gives the Planck distribution for the equilibrium energy spectral density of the solitons in some approximation.\cite{Chernitskii2018a}

\section{Conclusions}
\label{concl}
We have discussed the concept of unified field theory.
We have considered two nonlinear field models with world volume type action, namely extremal space-time film model
and  Born--Infeld nonlinear electrodynamics.

We have discussed the natural appearance of two long-range fundamental interactions, electromagnetism and gravitation, in these field models.

We have considered the quantum behaviour of the interacting solitons-particles.
In particular, we have introduced the concept of quasi-bounding quantization.

\end{document}